\documentclass[sigconf]{acmart}
\AtBeginDocument{%
	\providecommand\BibTeX{{%
			\normalfont B\kern-0.5em{\scshape i\kern-0.25em b}\kern-0.8em\TeX}}}

\renewcommand\footnotetextcopyrightpermission[1]{} 
\begin{document}

\title{On Decentralization of Bitcoin: An Asset Perspective}

\author{Ling Cheng}
\affiliation{%
\institution{Singapore Management University, Singapore}}
\email{lingcheng.2020@smu.edu.sg}

\author{Feida Zhu}
\affiliation{%
\institution{Singapore Management University, Singapore}}
\email{fdzhu@smu.edu.sg}

\author{Huiwen Liu}
\affiliation{%
\institution{Singapore Management University, Singapore}}
\email{hwliu.2018@smu.edu.sg}

\author{Chunyan Miao}
\affiliation{%
\institution{Nanyang Technological University, Singapore}}
\email{ascymiao@ntu.edu.sg}


\begin{abstract}
	Since its advent in 2009, Bitcoin, a cryptography-enabled peer-to-peer digital payment system,  has been gaining increasing attention from both academia and industry. An effort designed to overcome a cluster of bottlenecks inherent in existing centralized financial systems, Bitcoin has always been championed by the crypto community as an example of the spirit of \emph{decentralization}. While the decentralized nature of Bitcoin's Proof-of-Work consensus algorithm has often been discussed in great detail,  no systematic study has so far been conducted to quantitatively measure the degree of decentralization of Bitcoin from an asset perspective --
    How decentralized is Bitcoin as a financial asset? 
	
	We present in this paper the first systematic investigation of the degree of decentralization for Bitcoin based on its entire transaction history. We proposed both static and dynamic analysis of Bitcoin transaction network with quantifiable decentralization measures developed based on network analysis and market efficiency study.  Case studies are also conducted to demonstrate the effectiveness of our proposed metrics.
\end{abstract}

\maketitle

\section{introduction}
\label{sec:intro}

Since the publication of its white paper titled "A Peer-to-Peer Electronic Payment System" in 2008, Bitcoin (BTC) \cite{16_} has grown from a experiment among a coterie of cryptography enthusiasts into a distributed system running on a network of over 10,000 nodes globally, trading at a price of 40,000 US dollars each as of Feb. 7 2021. Simply put, Bitcoin is a digital currency with peer-to-peer transactions enabled not by traditional centralized institutions, but instead by a network of nodes through cryptography and recorded in a public distributed ledger called a blockchain. The fact that it has demonstrated for the first time that a financial asset with the fastest growing value ever can be maintained by a distributed network of computing nodes with no central authority in-charge, has established Bitcoin often as the best application of distributed ledger technology embodying the spirit of \emph{decentralization} 


While the notion of decentralization has been frequently discussed on the Proof-of-Work consensus algorithm underlying Bitcoin, very little research effort has so far been invested to assess the degree of decentralization of Bitcoin as a financial asset. Measuring decentralization of Bitcoin from a financial asset perspective is important due to two reasons. Firstly, Bitcoin, initially designed as a currency and more recently treated increasingly as an asset, needs to be thoroughly understood first and foremost by its financial aspects. Secondly, no clear relationship in terms of the degree of decentralization has been established between the financial aspects and the underlying consensus algorithm of Bitcoin.  


We therefore conduct in this work a comprehensive study of the decentralization of Bitcoin as an asset from two aspects -- 
(1) Asset At Rest. We examine the static distribution and other statistics of Bitcoin tokens at regular snapshots along its entire history, and (2) Asset In Motion.  We analyze the flow and other dynamics of Bitcoin tokens by its transaction network. 

	
It is important to point out that, the individual identity in Bitcoin is defined at address level, together with the anonymity and autonomy of each address in transaction by design \cite{16_}.  As our work is aimed to unveil the nature of and relation between transacting entities in the virtual realm of Bitcoin, \textbf{all results in our study are therefore based on address analysis}. While an interesting question, it is beyond the scope of this work to quest into the underlying real-world human beings mapped to each address. On the other hand, the highly complicated and diverse nature of the addresses (e.g, exchanges, ICOs, funds wallet, etc.) poses a practical challenge to the validity and usefulness of attempts trying to match each address to a unique real-world person or group. 


We identify for our work the following contributions. 
\begin{itemize}
    \item 
    We based our study on the entire history of Bitcoin from its very first transaction  with a daily granularity for an insight into the fullest picture of Bitcoin as an asset. For example, we observed that the whole lifetime history of BTC can be divided into three major stages with distinctive patterns. 
    \item
    We have examined the decentralization of Bitcoin's asset aspects from both static and dynamic perspectives. Analysis are given both to capture the degree of decentralization of Bitcoin token at important snapshots along timeline and to understand how the tokens have flown through the transaction network to connect observations at static snapshots. 
    \item 
    We have in all our analysis focused on quantitative measurement of the decentralization of Bitcoin. Instead of indiscriminately adopting network measures on the transaction network, we proposed carefully-chosen metrics based on their effectiveness of reflecting the decentralization of Bitcoin in the context of a financial asset: (1) Based on the dispersion of network metrics of Degree Centrality and  PageRank of top-100 token addresses; and (2) Based on Herfindahl-Hirschman Index(HHI), which is widely used in economics to measure market concentration.  We further illustrate these findings by detailed case studies to align with real-life events in Bitcoin community, lending interpretability to our data analytical results.
   
\end{itemize}

The rest of the paper is organized as follows.
We first describe our Bitcoin transaction data in Section \ref{sec:data_collect}. 
%
%
In Section \ref{sec:static_analysis}, we conducted a static analysis of Bitcoin's top addresses.
In Section \ref{sec:dynamic_analysis}, we conducted a dynamic analysis of Bitcoin based on its transaction graph.
We then discuss related work to give more background in Section \ref{sec:related}.
Finally, we conclude the paper in Section \ref{sec:conclusion}.


\section{Data}
\label{sec:data_collect}

\begin{figure}
	\vspace{3ex}
	\includegraphics[width=1\columnwidth, angle=0]{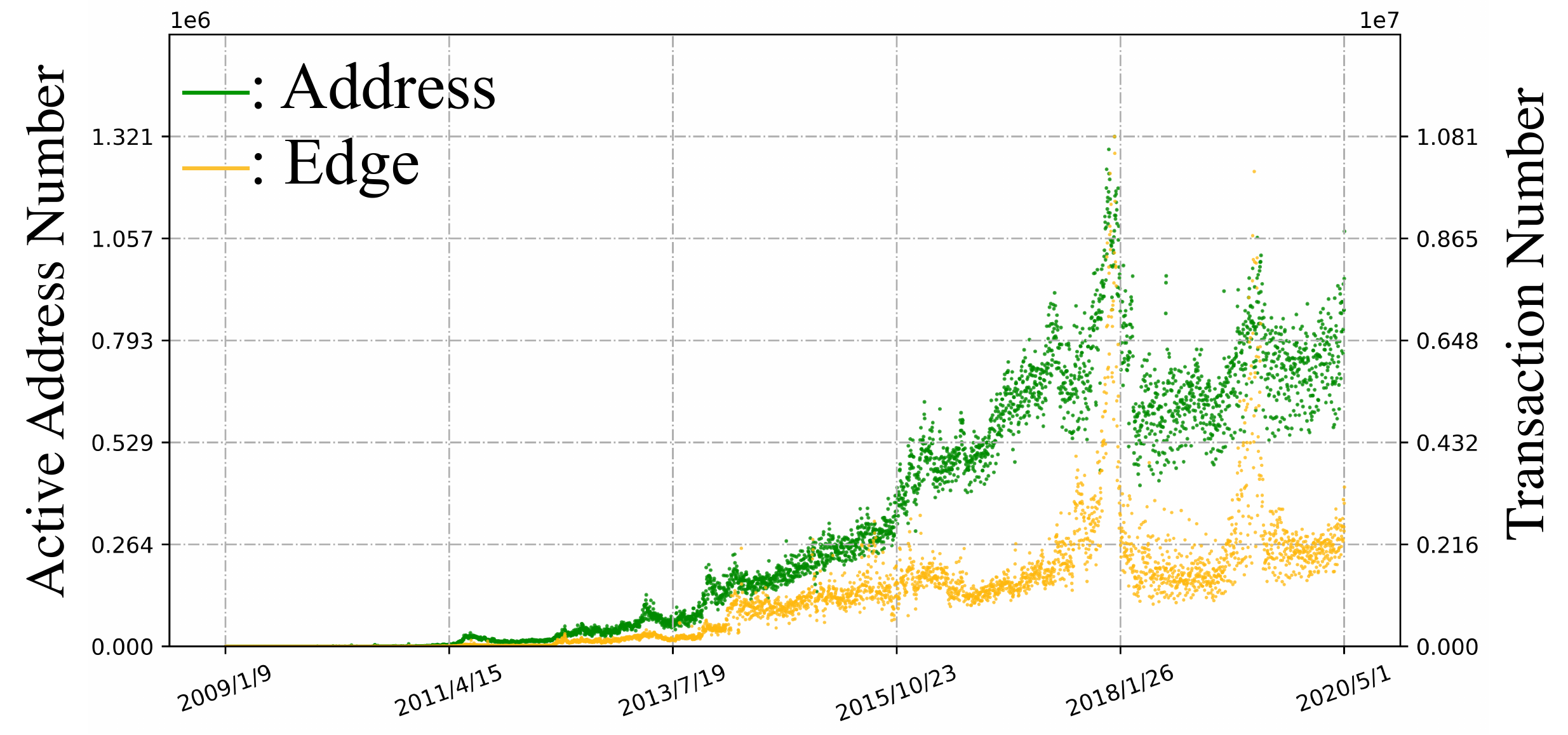}
	\vspace{-4ex}
	\caption{The history of daily active address number and transaction number} 
	\label{fig:histroy_overview}
	\vspace{-4ex}
\end{figure}

Thanks to its underlying distributed open ledger technology, Bitcoin transaction data is publicly available. Our Bitcoin transaction data\cite{15_} is from Jan 4th, 2009 to May 2nd, 2020. Daily transaction graph\cite{24_} is built with the balance of all addresses calculated and saved. 
Since in a Bitcoin transaction, the input-output address mappings are not explicitly recorded, we follow the processing as in\cite{10_}. A transaction with inputs from $N$ distinct addresses and outputs to $M$ distinct addresses is processed to $N\times{M}$ directed edges.

As shown in Figure \ref{fig:histroy_overview}, 
both the number of active daily addresses and number of daily transactions have demonstrated remarkable growth from a trivial number to $7.91\times{10^5}$ and $2.16\times{10^6}$ respectively. 
 
\section{Asset Static Analysis}
\label{sec:static_analysis}

\begin{figure}
	\vspace{-0ex}
	\includegraphics[width=1\columnwidth, angle=0]{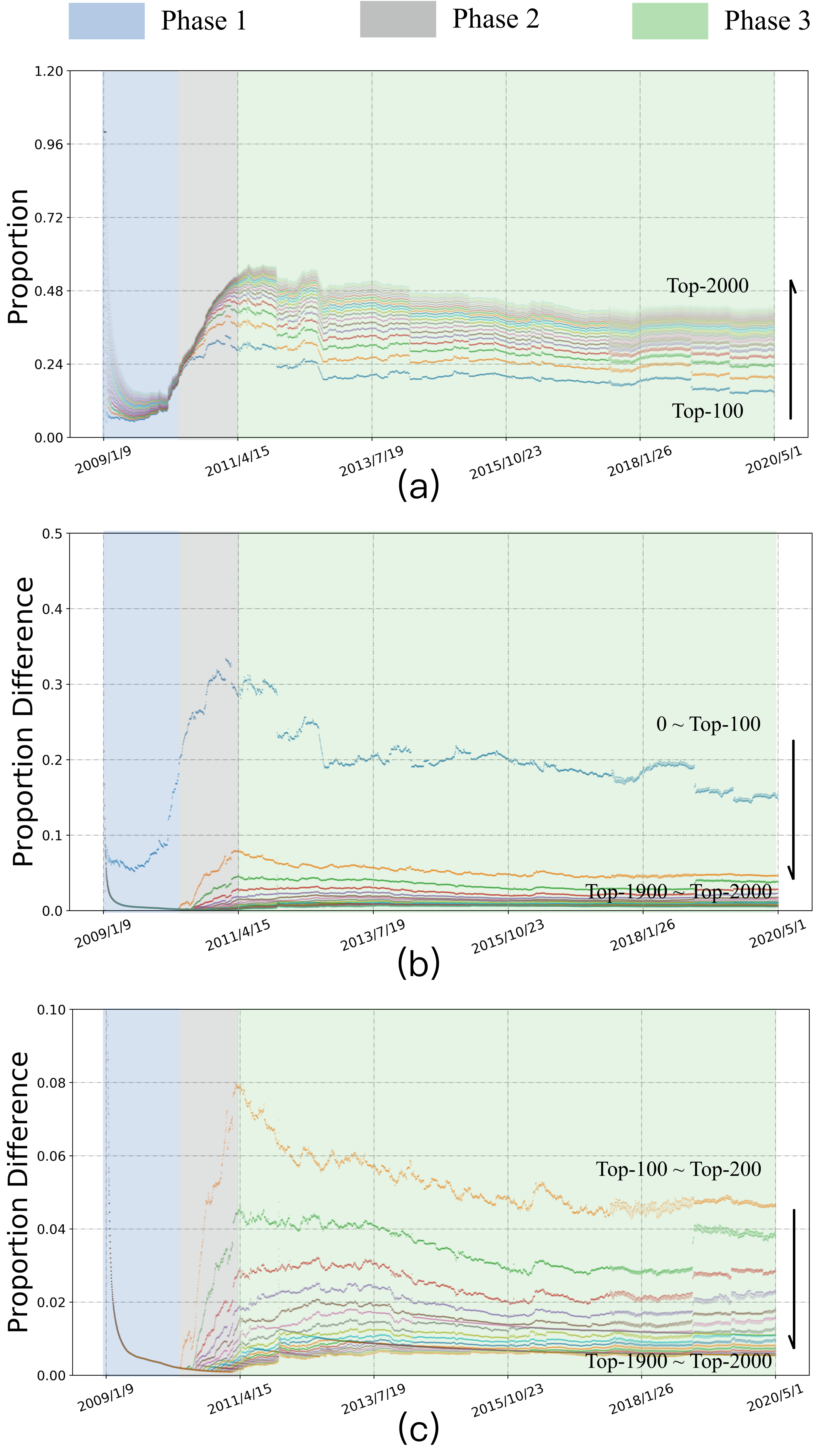}
	\vspace{-4ex}
	\caption{(a): Proportion of the Top-$x$ to the number of total bitcoins where $x\in(100,2000)$.
	Top-$100$ is about 0.424 of Top-$2000$.
	(b): Proportional difference between Top-$x$ and Top-$x$+$100$ where $x\in(0,1900)$. The proportion difference between Top-$0$ and Top-$100$ is equal to the proportion of Top-$100$. 
	(c): Zoom into the proportional difference between Top-$x$ and Top-$x$+$100$ where $x\in(100, 1900)$.}
	\label{fig:balance_proportion}
	\vspace{-3ex}
\end{figure}

\begin{figure*}
	\vspace{-0ex}
	\includegraphics[width=2.1\columnwidth, angle=0]{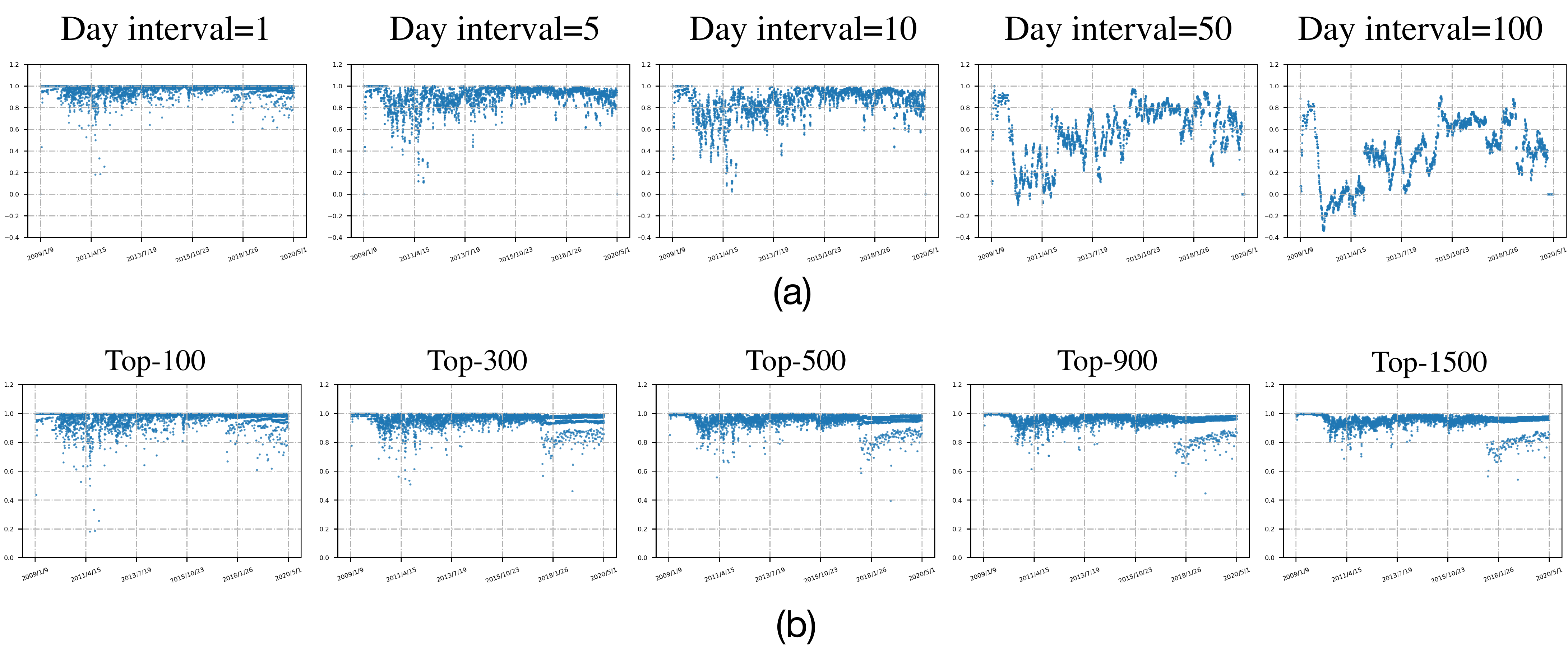}
	\vspace{-2ex}
	\caption{
    (a): Top-$100$'s Spearman Coefficient between the ranking of $i$ th day and $i+interval$ th day(interval=$1$,$5$,$10$,$50$,$100$).
    (b): Spearman Coefficient of top-x(100, 300, 500, 900, 1500) ranking with day interval = 1.}
    \label{fig:spearman_coefficient}
	\vspace{0ex}
\end{figure*}

\begin{figure*}
	\vspace{-0ex}
	\includegraphics[width=2.1\columnwidth, angle=0]{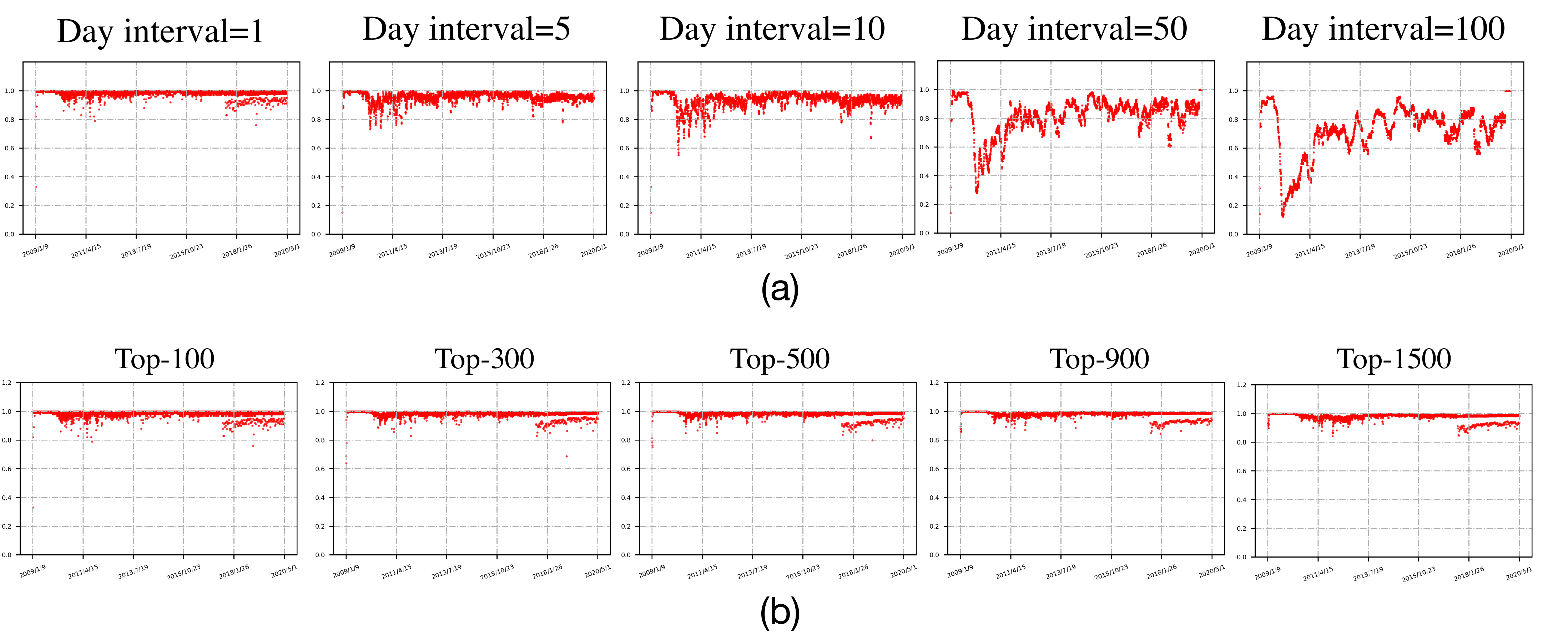}
	\vspace{-2ex}
	\caption{
    (a): Top-$100$'s Retention rate between $i$ th day and $i+interval$ th day(interval=$1$,$5$,$10$,$50$,$100$).
    (b): Retention rate of top-x(100, 300, 500, 900, 1500) list with day interval = 1.}
	\label{fig:retention_rate}
	\vspace{-2ex}
\end{figure*}

\begin{figure*}
	\vspace{-0ex}
	\includegraphics[width=2.1\columnwidth, angle=0]{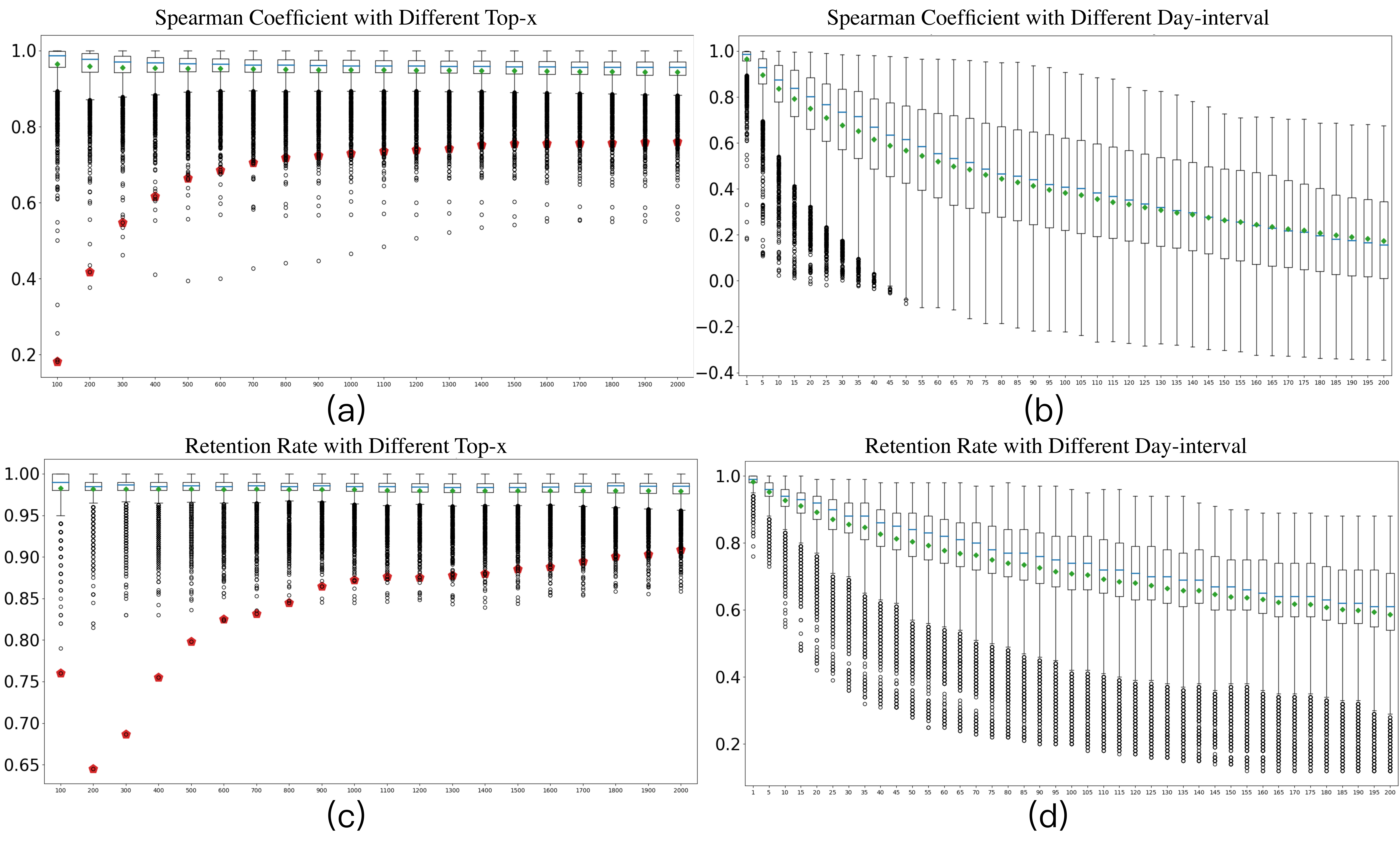}
	\vspace{-2ex}
	\caption{(a): Spearman coefficient of different Top-$x$ where $x\in(100,2000)$ (Red points are the data on 20 Jun, 2011),
	(b): Spearman coefficient of Top-$100$ with different  day interval,
	(c): Retention rate of different Top-$x$ where $x\in(100,2000)$ (Red points are the data on 3 Dec, 2018),
	(d): Retention rate of Top-$100$ with different  day interval.}
	\label{fig:boxplot}
	\vspace{-2ex}
\end{figure*}

\begin{figure}
	\vspace{-0ex}
	\includegraphics[width=1.\columnwidth, angle=0]{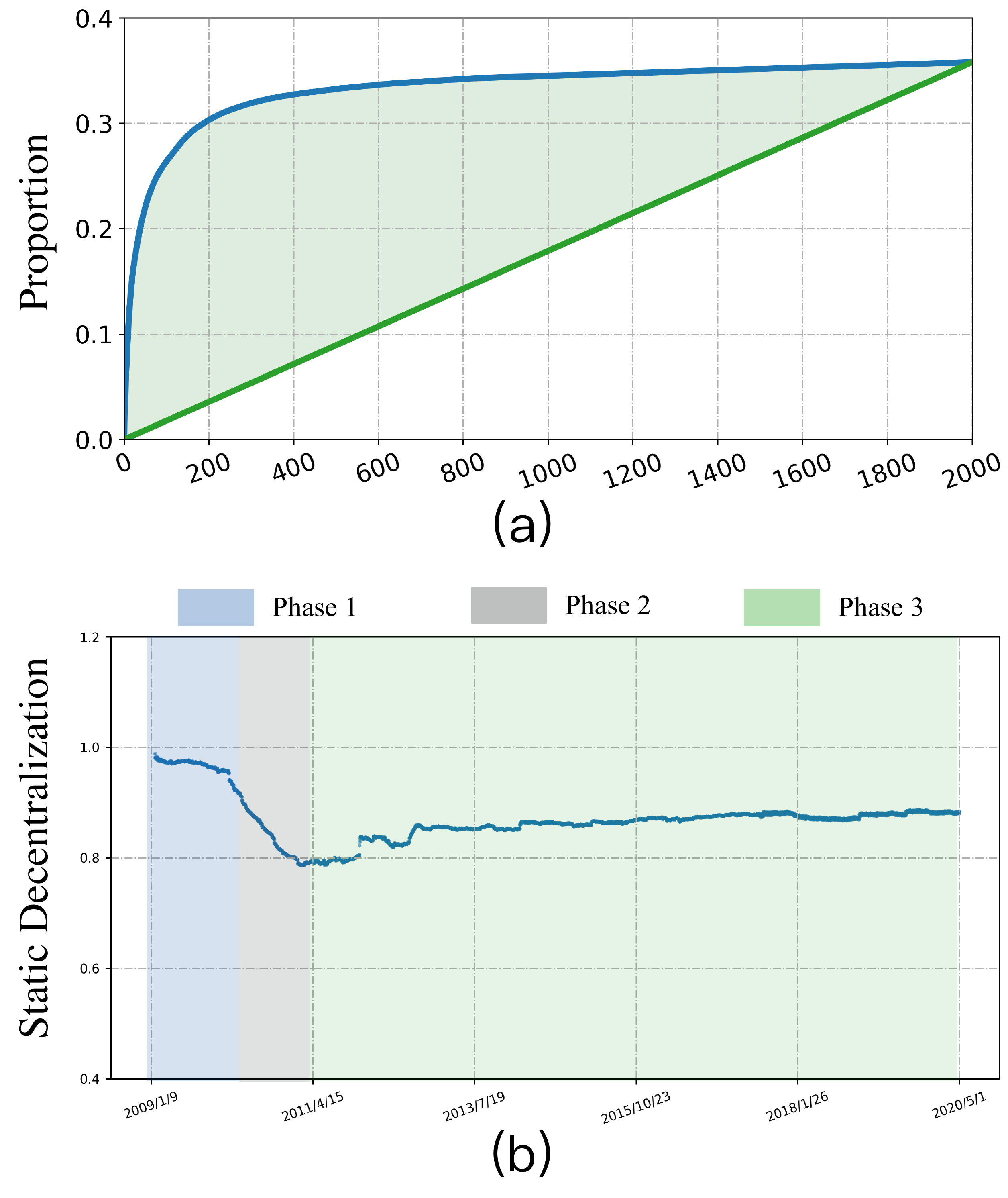}
	\vspace{-4ex}
	\caption{(a): Static decentralization degree definition. 
	The blue and the green lines stand for the Top-$x$ ($1-2000$) accumulative wealth proportion of real status and the status under equally distribution separately. 
	The shadow area is used to calculate the static decentralization degree. This image is based on the data of Sep 1st, 2010. (b): Full-history static decentralization degree.}
	\label{fig:static_decentralization}
	\vspace{-2ex}
\end{figure}

It is customary to measure the centralization of traditional financial assets by identifying the so-called list of wealthiest, e.g., Forbes Richest, and calculate how much of the total fortune are owned by these top-ranking people on the list. According to a recent study, world's 26 richest people own as much as the poorest $50\%$ of the world's entire population.\footnote{https://www.theguardian.com/business/2019/jan/21/world-26-richest-people-own-as-much-as-poorest-50-per-cent-oxfam-report} 
We refer to these types of study as \emph{static analysis} as they look essentially at snapshots of the asset distribution at certain timestamps.  

Likewise for Bitcoin, we conduct a static analysis of its degree of decentralization by investigating the following questions and trending the answers against its entire history.

 \begin{itemize}
    \item Which nodes are the top-ranking ones by BTC ownership?  
    
    \item Whether a relatively small percentage of nodes on this ranking list control a significant portion of the total BTC asset? 

    \item How does the degree of decentralization trend along time? 
    
    \item How do nodes in the top-ranking list change along time? 
\end{itemize}

\subsection{Ranking Analysis}
\label{sec:static:proportion_evolution}

We start by identifying the top-$2000$ addresses by BTC balances on a daily basis, i.e., the 2000 addresses that own the most BTC by day, for the entire Bitcoin history.   More important than the absolute amount of BTC is to understand the \emph{proportion} of the total BTC asset these 2000 addresses collectively command.  Furthermore, we are interested in finding out if there exists any subgroup among these 2000 addresses that dominates the centralization of BTC. As such a subgroup, if it exists, would be a better group to focus our study on than the  top-ranking 2000.  

We calculated the proportion of the total BTC asset commanded by the top-$N$ addresses for a sequence of different $N$s from $100$ to $2000$ with an increment of $100$ along time, as shown in the Figure \ref{fig:balance_proportion} (a). Furthermore, we zoom into each pair of top-$N$ and top-$(N+100)$ to calculate the difference in their BTC proportions among the total, and compare these differences across all pairs as $N$ ranges from $100$ to $1900$ as shown in Figure \ref{fig:balance_proportion} (b).  The study presents the following findings. 

\subsubsection{A Three-Phased History}
Firstly, the entire Bitcoin history can be identified with three distinct phases by the following two criteria for measuring its decentralization: 
\begin{enumerate}
    \item C1: The proportion of BTC owned by top-2000 addresses against the total BTC supply. This can be told by the y-axis value of the curves in Figure \ref{fig:balance_proportion} (a) -- The higher the curve, the greater the proportion and the lower the decentralization. 
    \item C2: The proportion of BTC owned by each top-$N$ group within the top-2000 addresses for $N$ ranging from 100 to 2000 with 100 increment. This can be told by the degree of separation of the curves in Figure \ref{fig:balance_proportion} (b) and (c) -- The more separated the curves, the greater the proportion differences for each top-$N$ and top-$N+100$ pair as $N$ increments and the lower the decentralization. 
\end{enumerate}

\vspace{2mm}
\noindent\textbf{Phase 1: Increasing decentralization by both C1 and C2}   (from Jan 9, 2009 to Feb 9, 2010) As this is the very initial phase of Bitcoin from its birth with the growth of participating nodes in the Bitcoin network. Almost all the curves exhibit a monotonous downward trend, a clear sign for increasing decentralization by C1. The dramatic steepness, driving the proportion of the top-2000 all the way down to a low point of 0.137, demonstrates the great momentum of both Bitcoin's growth and its decentralization. The fact that all curves in Figure \ref{fig:balance_proportion} (c) merges into one demonstrates the evenness in BTC distribution within the top-2000 addresses, except for the top-100 group which is illustrated separately in \ref{fig:balance_proportion} (b), serving as another clear sign for increasing decentralization by C2. 

It is also worth noting from Figure \ref{fig:balance_proportion} (a) that the higher-ranked the node, the more stable its proportion. In particular, the bottom blue curve (the top-100 nodes) is relatively flat around 0.049.
It is fair to say that, during this phase, there is a continuous increase in the decentralization of Bitcoin while the top-100 nodes have a fairly steady control over roughly $5\%$ of the total BTC. 

\vspace{2mm}
\noindent\textbf{Phase 2: Decreasing decentralization by both C1 and C2} (from Feb 10, 2010 to Mar 23, 2011): This phase witnesses a sharp decrease of decentralization both within the top-2000 addresses and against the total BTC asset, as shown by the simultaneous separating and rising of all the curves throughout this phase in Figure \ref{fig:balance_proportion} (c), serving as a clear sign for decreasing decentralization by both C1 and C2. 
The end of this phase is marked by a simultaneous climax for both C1 and C2, as illustrated by the peak of all the curves in Figure \ref{fig:balance_proportion} (a) and the maximum separation among all the curves in Figure \ref{fig:balance_proportion} (c), both of which are followed by a gradual decline heralding the beginning of Phase 3. 

\vspace{2mm}
\noindent\textbf{Phase 3: Stabilizing decentralization with a gradual increasing trend by both C1 and C2}  (from Mar 24, 2011 to the present day): This ongoing phase illustrates a stabilizing and slowly increasing degree of decentralization of Bitcoin, with the top-2000 addresses declining from $60\%$ toward $40\%$ and  the top-100 group, a half of that, from $30\%$ to $20\%$. Contrasted with Phase 2 during which all curves spread out from a merged single, Phase 3 sees the curves flatten out in an almost parallel fashion with fluctuations over time. It can be concluded that, the degree of decentralization of Bitcoin has largely stabilized after a slow yet steady trend of increase since Phase 2. 

\subsubsection{The Importance of Top-100}
It is observed from Figure \ref{fig:balance_proportion} (a) that, from Phase 2 onward,  the top-100 nodes consistently control around $40\%$ of all the BTC controlled by the top-2000 list. On the other hand, the overwhelmingly-larger gap between the top curve (representing the proportion of top-100) and the second curve (representing the proportion difference between top-200 and top-100) in Figure \ref{fig:balance_proportion} (b) illustrates the dominance of top-100 group among the top-2000. It follows that, in terms of studying the decentralization of the top-2000 nodes trending against time, we can effectively focus on the top-100 nodes as a proxy. This is why in the dynamic analysis in Section \ref{sec:dynamic_analysis}, we would focus on the top-100 list.

\subsection{Ranking Stability}
\label{sec:static:ranking_stability}
In Subsection \ref{sec:static:proportion_evolution}, we examined the degree of decentralization of BTC by examining the proportion of BTC owned by top-ranking addresses, i.e., the asset concentration level.  Another aspect of the degree of decentralization is to evaluate how perpetuated the membership and order of the top-ranking lists are along time, which we call ranking stability. The more stable the ranking, the lower the degree of decentralization. 

To investigate how nodes in the top-ranking list change along time, we adopt two complementary measures: (I) Spearman coefficient; and (II) Retention rate.  Spearman coefficient, which is a classic measure to estimate the correlation between two variables $X$ and $Y$, is used to examine the extent to which nodes change order in the ranking as an indicator of ranking stability.   For two rankings on day $i$ and day $i+n$ respectively, which we call a $n$-day interval, we take the ranking on day $i$ as variable $X$ and that on day $i+n$ as $Y$, and calculate the spearman coefficient for $X$ and $Y$. The closer to 1 the absolute value of the spearman coefficient, the greater the consistency between the two rankings, and hence the greater the ranking stability.  On the other hand, the closer to 0 the absolute value of the spearman index, the lower the consistency between the two rankings, and hence the less the ranking stability.  Retention rate is used to find out how much membership replacement take places in the ranking of top-$N$ lists. It is calculated as the ratio of the number of existing addresses to $N$, the size of the ranking list.  Note that the two measures complement each other to offer a more comprehensive evaluation of ranking stability, e.g., addresses in a ranking list with no new address added could still drastically change their relative ranking orders. 
%
%
%
Studies on the ranking stability reveal the following findings:

Firstly, the ranking stability of top-$N$ addresses for one day interval are largely consistent over time for $N$ from 200 to 2000, as illustrated in Figure \ref{fig:spearman_coefficient} and Figure \ref{fig:retention_rate}. From Figure \ref{fig:boxplot} (a), it can be observed that the spearman coefficient for one day interval for $N$ from 200 to 2000 gradually converges to a mean of 0.958 and standard deviation of 0.042. 
From Figure \ref{fig:boxplot} (b), it can be observed that the retention rate for one day interval for $N$ from 200 to 2000 gradually converges to a mean of 0.985 and standard deviation of 0.013.

Secondly, the ranking stability of top-100 addresses distinguish itself from the rest, as illustrated in Figure \ref{fig:spearman_coefficient} and Figure \ref{fig:retention_rate}. From Figure \ref{fig:boxplot} (a), a significant increase in  interquartile range (IQR) can be observed from top-100 to top-200, which indicates a much greater variance in ranking stability across different days for top-200 list when compared against top-100.  It follows that a noticeable drop in randomness is observed in ranking stability for top-100, a phenomenon made more pronounced when one notices that the smooth and gradually reducing IQR for the boxplots from top-200 all the way to top-2000.  Combined with the observation that the drop in spearman coefficient from top-100 to top-200 is also the greatest among all, with all the other spearman coefficients sharing almost the same value, it leads to the result that top-100 distinguishes from the rest by commanding a ranking stability that is both much higher and much less random. 

Furthermore, from the retention rate in \ref{fig:retention_rate}, we find that the variance in retention rate for top-100 is the greatest among all, illustrated by the much greater IQR of the boxplot.  When put together this finding and the previous result on the ranking stability, we can conclude that there exists in top-100 a core set of addresses that enjoys a high level of stability for both their membership and ranking order in the top-100 list, a negative indicator of high decentralization which will also be echoed by our dynamic analysis in Section \ref{sec:dynamic_analysis}. 

Thirdly, when we increase the day interval $N$ from 1 to much larger values, top-100 list displays continuously lower ranking stability and retention rate, as observed from Figure \ref{fig:spearman_coefficient} and Figure \ref{fig:retention_rate}. This means, from a temporal point of view, even top-100 addresses, the most stable group in BTC ecosystem, exhibit decentralization in the sense that new addresses would consistently replace a significant portion of the existing ones over a sufficiently-long period of time (e.g., $40\%$ in 200 days), rather than recurring groups of addresses taking turns to monopolize the top-100 list.

\subsection{Case Study}
\label{sec:static:case study from boxplot}
To better demonstrate and relate the spearman coefficient to real-life scenario behind the data, we zoom into data points of particular interest and present two case studies.  Now that we know that top-100 is the most stable group in terms of ranking, one might wonder what is the story behind some of the days that give the lowest values for the spearman coefficient in Figure \ref{fig:boxplot} (a). 

\vspace{2mm}
\noindent\textbf{June 6th 2011.}
For the first case, we focus on the day corresponding to the lowest point for top-100, trace out this data point in the plot of top-$N$ for all $N$ from 200 to 2000, and color them red, as shown in Figure \ref{fig:boxplot} (a).

These red points are the data of June 6th 2011.  On that day Mt Gox, a bitcoin exchange based in Japan, got hacked and a large mount of BTCs were stolen from Mt Gox.  As Mt Gox, through multiple addresses under its control, was handling over 70\% of all BTC transactions worldwide as the largest BTC intermediary and the world's leading BTC exchange at that time, it is not hard to see why the ranking in the top-100 list changed so much as a result of that incident. 
It is also interesting to see why that day did not cause a ranking change as drastic for larger $N$(e.g, $N > 1000$ as it did for top-100. It is highly likely that the BTCs were transferred to addresses still ranked among the top-2000.

\vspace{2mm}
\noindent\textbf{December 3rd 2018.}
For the second case study, we refer to the red-colored points in Figure \ref{fig:boxplot} (c). For the Retention rate, 
these red points are the data of December 3rd 2018, on which day there was a significant drop in Bitcoin mining difficulty. Consequently, miners were rewarded with much more BTCs from mining all of a sudden.  It is understandable that this could lead to a surge of new addresses entering the top-ranking lists.  What is the most interesting though, is that the group with the greatest change is top-200 while the retention rate of that day no longer saw big changes as $N$ increases.  It can be thus inferred that most of these benefiting miners were ranked within top-1000 before the mining difficulty drop and, while they caused significant changes to most groups for $N < 900$ (they represent the lowest data points), they caused the most update in list membership for group top-200. In comparison, not as many successfully entered the top-100 list.

\subsection{Static Decentralization Degree}
\label{sec:static:static_decentralization_degree}
After we examine the top-ranking lists and their stability, it is also important to study the overall decentralization degree of all the addresses as a whole, which we call the \emph{static decentralization degree}.  We still use the top 2000 addresses as they represent the majority of the asset and activity in the BTC ecosystem. 
We design our static decentralization degree with the idea that the maximum decentralization in terms of static asset distribution is when everyone owns the same amount of asset in the system. This situation is depicted in Figure\ref{fig:static_decentralization} (a) by the green straight line, which represents the cumulative proportion of BTCs owned by top-$N$ addresses as $N$ increases.  The blue curve representing the real situation would be a convex curve deviating upward from this green straight line by definition.  The more bulging the curve, the more uneven the distribution of BTCs among the top 2000 addresses, and accordingly the lower the decentralization. We therefore define the static decentralization degree as follows.

we measure the area between the curve of cumulative proportion of the top-$N$ addresses and the straight line of equal distribution,
and the static decentralization degree, denoted by $D_{static}$, is defined by 1 minus this area. Specifically, we have 

\[
 D_{static} = 1-\int_0^{N} \left(\mathbb{C}_r(x) - \mathbb{C}_e(x)\right) dx
\]
where $\mathbb{C}_r(x)$ and $\mathbb{C}_e(x)$ represents respectively the cumulative proportion of BTCs owned by top-$x$ addresses in the real case and in the ideal case where every address owns the same amount.

The blue curve in Figure\ref{fig:static_decentralization} (a) represents the data on September 1st, 2020. The sharp rise of the curve before $x=200$ and the flatness thereafter, as well as the corresponding bulging of the curve toward $x<200$ indicate the high centralization of BTC asset in the top-200 addresses by that time, echoing the results in ranking analysis in Subsection \ref{sec:static:proportion_evolution}.

We can have a clearer overview of the static decentralization degree for the full history of BTC as shown in Figure\ref{fig:static_decentralization} (b). Interestingly, if we show the static decentralization degree curve against the three phases as in Subsection \ref{sec:static:proportion_evolution}, the distinct patterns exhibited by different segments of the curve align well with the boundaries of the three phases, a strong support for our definition, and characterizations of decentralization degree, of the three phases from another perspective. 

\section{Asset Dynamic Analysis}
\label{sec:dynamic_analysis}

As an asset, we are interested not only in its static distribution at any given timestamp, but also in its dynamic flow through transactions. In Section \ref{sec:static_analysis}, we have examined the level of decentralization of BTC from a static timstamp point of view. In this section, we will continue to evaluate its level of decentralization from a dynamic perspective by studying its transaction network, i.e., whether, and to what extent, the BTC transactions are monopolized by a certain subgroup of nodes. 

In previous analysis, we have demonstrated the distinct importance of top-100 addresses in BTC ecosystem and established the validity of focusing on this group as a surrogate for the whole. We therefore will zoom onto the transaction network centering around top-100 addresses, i.e., all transactions involving at least one address from the top-100 group.

In Subsection \ref{sec:graph_proper_analysis}, we will study the level of decentralization of BTC transaction network by a set of network properties and their dispersion. In Subsection \ref{sec:market_efficiency_analysis}, we adopt a perspective of market efficiency, which takes a deeper root in the finance community, to examine the same level of decentralization of BTC transaction network.

\subsection{Decentralization by Network Properties}
\label{sec:graph_proper_analysis}

When BTC transactions are represented in the format of a network, it is natural to ask how network properties can be used to offer the most insight in measuring dynamic decentralization from transactions.  To that end, we adopt a two-step approach as follows.

Firstly, we identify network measures that strongly indicate the node's importance in controlling transactions in the network. In network analysis, the following  three network metrics --- Degree Centrality (In-degree and Out-degree) and PageRank --- come the closest to what we aim to characterize. 

\vspace{2mm}
\noindent\textbf{Degree Centrality.} In a transaction network, in-degree and out-degree carry different meaning and significance, just like in the case of hyper-links between webpages. In-degree means asset transferring into the target node and therefore naturally indicates greater trust and endorsement for the target node than out-degree. However, as we observed that the In-degree and Out-degree have very similar distribution in our study, we use Degree Centrality to summarize both. In general, the greater the Degree Centrality, the more important the node due to involvement in more transactions. 

\vspace{2mm}
\noindent\textbf{PageRank.} The reason why PageRank, the classic network measure for ranking the authority of web pages, is relevant in our setting is that -- Transactions can be viewed as trust endorsement. The
higher the PageRank score of a node, the more trust endorsement for the node as manifested by BTC transactions transferring the asset to this node\cite{33_, 34_, 35_}.
In general, a node with high PageRank score in a financial network is considered to serve as intermediary of three kinds: (I) financial intermediary, serving as middleman for financial transactions to create efficient markets and to lower business cost; (II) credibility intermediary, offering trust to other nodes to enable transactions; and (III) information intermediary, providing information for asset pricing, bidding, matching and others based on data from large transaction volume.

Secondly, it should be noted that the notion of decentralization does not necessarily depends on the importance in transaction control for any particular node as measured by any one network property, it is more related, rather, to the disparity among different nodes in their values for any chosen network property.  As such, it is not sufficient simply to identify the network properties, more importantly, we propose to use the notion of \emph{dispersion} and use it on each metric to evaluate the decentralization of the transaction networks. We define dispersion  $d_{m}$ for a chose metric $m$ as follows:
\begin{equation}
\label{eq:metric_dispersion_degree}
d_{m}={\frac{{H_{m}-L_{m}}}{{AVG_{m}-L_{m}}}}
\end{equation}
Where ${H_{m}}$, ${L_{m}}$ and ${AVG_{m}}$ stand for the greatest, the least and the average value for metric $m$.  The higher the dispersion, 
the more pronounced the control of transactions by a subgroup within the whole graph, and the less the dynamic decentralization of BTC transaction network. 

To better understand dispersion in our setting, we explore a few cases for which the dispersion takes some specific values. Note that, by definition,
for a chosen metric, if the largest value is much greater than the second largest value, the dispersion would take a value close to 100.
If the largest and the second largest values are close, and these two values are much greater than the third largest value, the dispersion would take a value close to 50. It is not hard to extend similar analysis to other cases. It is therefore possible to use dispersion to estimate the number of addresses that form the top-ranking group such that their metric values are close to one another and, as a whole, much greater than the rest.  
\begin{figure}
	\vspace{-0ex}
	\includegraphics[width=1\columnwidth, angle=0]{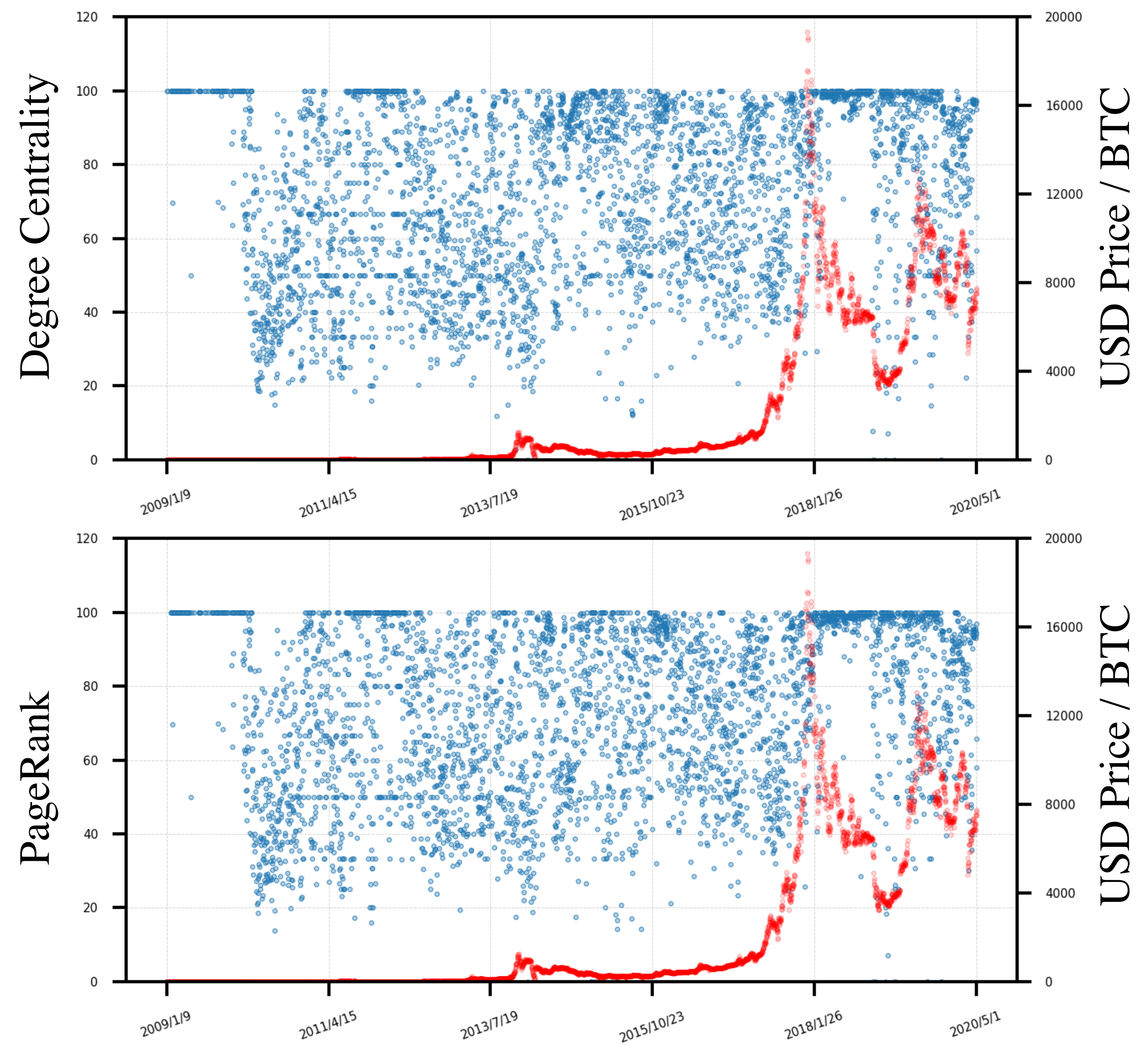}
	\vspace{-3ex}
	\caption{Dispersion of Degree Centrality, PageRank. The red line stands for bitcoin price.}
	\label{fig:metric_and_dispersion_degree_of_metric}
	\vspace{-2ex}
\end{figure}

As we observe these three metrics, we have the following findings. First of all, it is observed that for most of the time, the dispersion are above 20,
which means only fewer than 5 addresses were in uniquely important positions in terms of controlling transactions in the network.

\subsection{Decentralization by Market Efficiency}
\label{sec:market_efficiency_analysis}
While designed as an electronic payment system, Bitcoin has evolved to be a much more complicated combination of currency, commodity, security and even asset. The transactions of BTC can therefore viewed indeed as a financial market. It is therefore natural and useful to study the level of decentralization of this network from market efficiency perspective. 

The Herfindahl-Hirschman Index (HHI) is a classic measure of market concentration 
and has been widely used to evaluate market efficiency. The definition is given by:
\begin{equation}
\label{eq:HHI_calculation}
HHI = \sum_{i=1}^{n} 10000*({H_i})^{2}, \hspace{1mm} H_i = {\frac{h_i}{C}},
\end{equation}
Where $h_i$ is the BTC held by entity $i$, $C$ is the total BTC minted by that time. The higher the HHI, the more significant the market concentration, and the less the decentralization.  In general, a market with an HHI of less than 1,500 is considered to be a competitive market, an HHI of 1,500 to 2,500 a moderately concentrated market, and an HHI of 2,500 or greater a highly concentrated market\cite{36_, 37_}. \footnote{https://www.investopedia.com/terms/h/hhi.asp}

The original definition of HHI comes with a concept of firm, which in our setting will be interpreted as an entity that can be considered as a coordinated body of nodes for the control of BTC. In reality,  the anonymity nature of BTC means we cannot determine whether a set of addresses belong to a same person or entity. We therefore propose three approaches (A1 to A3) to cluster addresses into entities that can be considered as firms in HHI.
\begin{enumerate}
	\item \textbf{A1:} Each address will be treated as a distinct firm.
	\item \textbf{A2:} We build the full-history transaction graph which contains only transactions between top-100 addresses. With community detection, addresses will be clustered into coordinated entities as firms. All top-100 addresses with no transaction edges will be each treated as a single firm. 
	\item \textbf{A3:} We build the full-history transaction graph which contains all transactions involving top-100 addresses. With community detection, addresses will be clustered into coordinated entities as firms. Coinbase(the system address issuing BTC) will be treated as a single node  $V_c$ and all addresses not in top-100 will be grouped as another single node $V_o$. 
\end{enumerate}

\begin{figure}
	\vspace{-0ex}
	\includegraphics[width=1.\columnwidth, angle=0, scale=1.0]{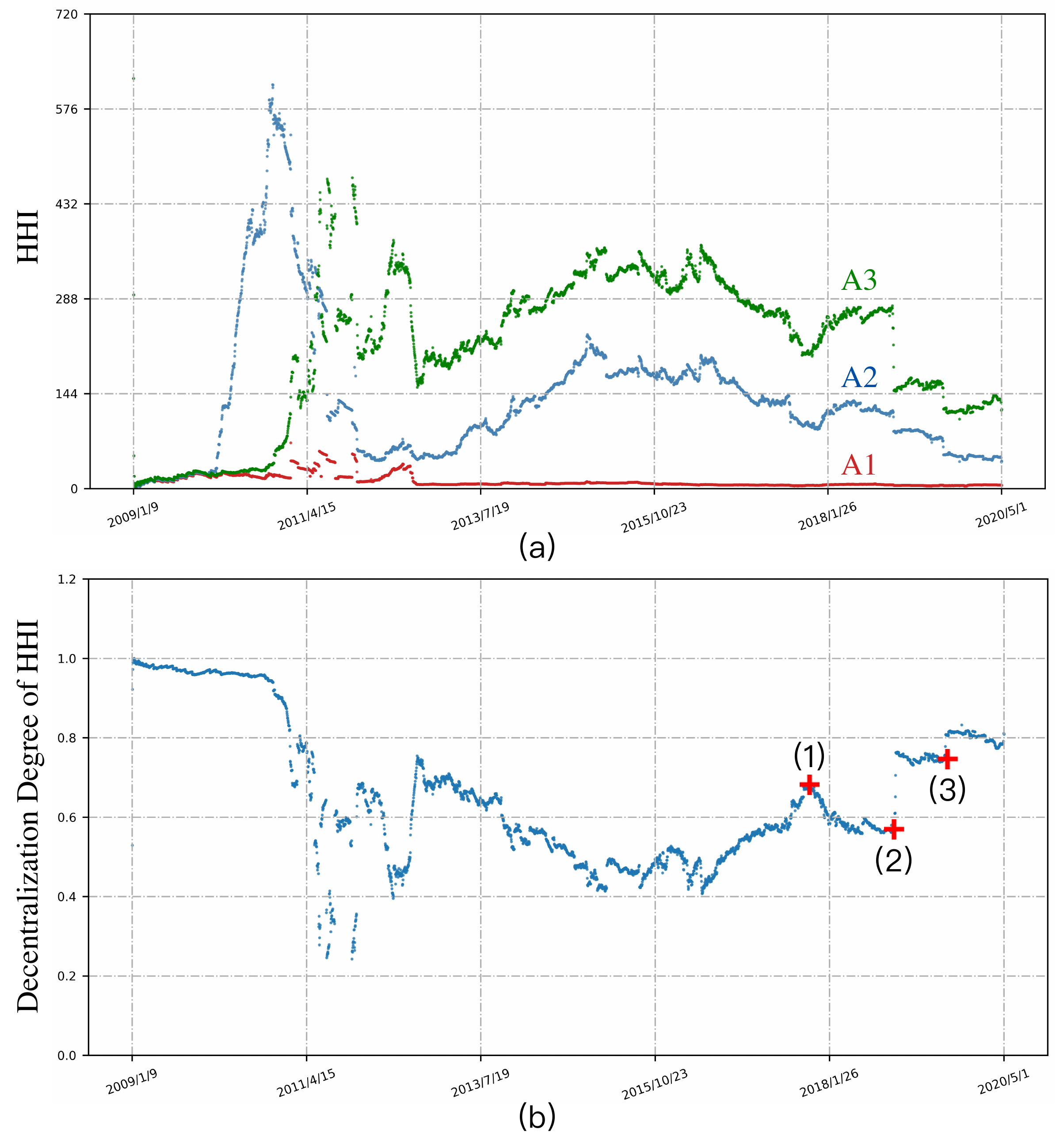}
	\vspace{-4ex}
	\caption{(a): HHI results of three different assignment methods. The red, blue and green lines stand for A1, A2 and A3 clustering approach correspondingly.
	(b) Decentralization degree based on HHI. Point 1, 2 and 3 stand for the data on October 25th 2017, December 5th 2018 and August 1st 2019 correspondingly.}
	\label{fig:HHI_analysis}
	\vspace{-2ex}
\end{figure}

Figure \ref{fig:HHI_analysis} (a) shows the HHI results calculated on all participating addresses (not just the top-100) at any given time stamp for the entire BTC history. Different colored curves represent the three clustering approaches respectively. We have the following findings.

First, compare A1 and A2, the largely flat-shaped A1 represents much lower HHI throughout the entire BTC history, which means, no single address has ever significantly dominated the market, indicating a high level of decentralization if all addresses are viewed as independent entities. 

Secondly, the sharp rise in the curve of A2 and A3 in the early phase of BTC history is attributed to the fact that the number of participants was relatively small back then and top-100 addresses controlled a majority portion of all BTCs, clustering of addresses from this group would necessarily result in increased HHI values.

Thirdly, compare A2 and A3, A3 essentially separate the group of miners and the group of addresses transacting with non-top-100 ones from the other top-100 addresses. 
It can be observed that (I)During the early days of BTC, 
the level of decentralization is relatively high with no dominating entities and perpetuated structures.  After considering Coinbase and non-top-100 addresses in the clustering as in A3, large groups in A2 are divided into small groups in A3, 
resulting in the drops of HHI values.  (II) The curves share similar growth patterns from 2012 onwards, which is due to the fact that, in later phases of BTC history, the top-100 group has formed a relatively stable structure with giant entities dominating the transactions such as large mining pools and major cryptocurrency exchanges.

We continue to propose a decentralization degree to quantify dynamic decentralization for BTC transaction network as follows based on A3, we think it is the most appropriate one to cluster addresses among the three approaches.  
\begin{equation}
\label{eq:metric_2}
D_{HHI} = 1 - normalized(HHI_{A3})
\end{equation}
where, $D_{HHI}$ is the decentralization degree based by market efficiency. $normalized()$ is the normalization function and $HHI_{A3}$ denotes the HHI calculated by clustering approach A3. The greater the value of $D_{HHI}$, the higher the level of decentralization. 

Figure \ref{fig:HHI_analysis} (b) shows the $D_{HHI}$ the decentralization degree based by market efficiency over the entire BTC history. One important difference can be observed between $D_{HHI}$ and $D_{static}$ the decentralization degree based on static BTC distribution analysis is that $D_{HHI}$ exhibit much greater fluctuation in later phases of BTC history. One may thus infer that the decentralization level of BTC  transaction network from the market efficiency perspective is much less stable. In particular, we zoom into three data points of drastic changes, as marked by (1), (2) and (3) in Figure \ref{fig:HHI_analysis} (b), to take a closer look at how they link to real-life events at that time.  

\vspace{2mm}
\noindent\textbf{(1) October 25, 2017:}
        Bitcoin Gold was launched. Contrary to what had been promised by the Bitcoin Gold pitch that
        Bitcoin Gold would bring greater decentralization, Bitcoin Gold price plunged over $66\%$ within the first couple of hours.
        Major market players therefore shifted their trust and confidence to Bitcoin instead as alternative and acquired significant amount of BTC in their portfolio, leading to a downward trend in the decentralization of BTC transaction network.

\vspace{2mm}
\noindent\textbf{(2) December 5, 2018:}
        BTC mining difficulty dropped significantly on that date, which gave rise to the public's worry for a potential $51\%$ attack. The damage to users' trust and confidence in Bitcoin had led to panic selling of BTC and caused the abrupt jump in the decentralization level. 
        
\vspace{2mm}
\noindent\textbf{(3) August 1, 2019:}
        Breaking news of massive data leakage from Binance, a top cryptocurrency exchange, was out on that date, sparking worry in public about the safety of BTC market in general. Like in case (2), the damage to users' trust and confidence in these large cryptocurrency exchanges led to wild selling which caused the abrupt jump in the decentralization level.


\section{related work}
\label{sec:related}

There have been research work proposed to investigate  decentralization of BTC: Some  focus on network congestion, delay, and other technical details to evaluate the performance of distributed network.\cite{1_, 2_, 28_, 18_, 19_, 21_, 29_} Others expose the less-than-optimized decentralization in terms of services  decision-making, transactions, and mining in the BTC system. \cite{3_, 27_, 17_, 20_, 22_, 32_}  These work, however, provide no quantitative evaluation of decentralization of Bitcoin. 

Examination of Bitcoin's decentralization from a quantitative perspective include \cite{5_} which proposes a centralization factor based on the ratio of uniformity of mining data. Their analysis is constrained to the mining reward 
and does not concern BTC transactions.
\cite{6_} focuses on evaluating the critical value of the number of nodes needed to control over 51\% of the network by using a Nakamoto coefficient.
This work analyzes the decentralization of the BTC network from the perspective of security in the consensus protocol.
\cite{7_, 8_, 9_} describe decentralization through the randomness degree of data. 
They use statistics and information theory, 
and calculate the variance coefficient and information entropy on blocks mined and address balance to quantify the decentralization of the BTC system.
%
%
Most above-mentioned work mainly focus on mining data and offers little insight for the transaction network.

Another line of work investigate the relation between BTC with financial activities. 
\cite{26_, 30_} present the relation between some financial malicious activities and blockchain technology and point out risks and regulatory issues as
Bitcoin interacts with conventional financial systems.
\cite{4_} analyzes the occurrence of large BTC transactions at certain points in time. 
\cite{10_, 11_, 31_} conduct detailed graph measurements on the transactions graph, without relating these results with decentralization though.
\cite{13_} has applied a heuristic for de-anonymization and extracted the user’s graph by merging addresses that belong to the same individuals. 
But with modern wallets and the advent of mixers (Mixing service 2019), the de-anonymization heuristic is no longer effective. 
%

\section{Conclusion}
\label{sec:conclusion}
	
In this paper, we provide the first study of the degree of decentralization for Bitcoin from a financial asset perspective based on its full history. We examine Bitcoin's decentralization from static analysis on the token distribution by daily snapshots over its whole history and dynamic analysis on Bitcoin's transaction network, a comprehensive and close look at Bitcoin's whole life cycle till the present day. 

Our studies reveal a set of interesting and consistent findings including a three-phased history for which distinct patterns of decentralization are observed to characterize each phase and a uniquely important top-100 address group, which is characterized by much greater ranking stability, higher degree of asset centralization and relatively stronger control over asset flow in the transaction network.  We also conduct dynamic analysis on Bitcoin's transaction network with both carefully-chosen network measures (i.e, degree centrality and PageRank) and the Herfindahl-Hirschman Index (HHI) from a financial market efficiency perspective. 
We also complement our statistical findings with case studies relating data phenomena of particular interest to real-life events in Bitcoin community to lend interpretability to data analysis results.


{\small
	\bibliographystyle{unsrtnat}
	\bibliography{main}
}

\end{document}